# An RFID-Based Assistive Glove to Help the Visually Impaired

Paniz Sedighi, Mohammad Hesam Norouzi, Mehdi Delrobaei, *Member, IEEE*
delrobaei@kntu.ac.ir - http://orcid.org/my-orcid?orcid=0000-0002-4188-6958

Faculty of Electrical Engineering, K. N. Toosi University of Technology, Tehran, Iran

Recent studies have focused on facilitating perception and outdoor navigation for people with blindness or some form of vision loss. However, a significant portion of these studies is centered around treatment and vision rehabilitation, leaving some immediate needs, such as interaction with the sur- rounding objects or recognizing colors and fine patterns without tactile feedback. This study targets such needs and delivers a straightforward communication method using a wearable, unobtrusive device with the environment. We initially discuss the advantages and limitations of related works to draw out the best-fitting design concepts. Then, we introduce the potential for emerging technologies such as radio-frequency identification. We present the design details and the experimental results of an assistive glove to allow people with vision disabilities to interact with the environment more efficiently. Based on the collected data from 17 blind-folded healthy participants, the implemented system's success rate in identifying objects was about 96.32%. Overall, 70% of the users found the device very satisfactory.

## I. Introduction

A statistical report recently released by WHO [1] shows that at least 2.2 billion people are struggling with vision impairment or blindness [2]. Besides low vision caused by old age, vision impairment can be congenital, mainly in the younger age groups.

Visually impaired are often challenged in their day-to-day tasks. They often lead a normal life; however, they face many troubles due to inaccessible infrastructure and social challenges. Moreover, the misunderstood perspective of society often leads to inconvenience.

Blind or visually impaired professionals also encounter barriers while interacting with objects and carrying out their job duties in office or industrial settings.

Assistive technologies are pivotal in enabling the visually impaired to perceive their surroundings and interact with objects. In a familiar environment such as a workplace or home where navigation is not the primary problem, low vision causes difficulties in interaction with the environment, such as discerning colors or identifying detailed shapes and patterns. Since traditional assistance methods such as the white cane could not provide a full spectrum of the environment through tactile information, inventions such as ALVU [3], QD laser [4], and Aira [5] came up with solutions to meet the requirements of people with visual disabilities. While the white cane is supplemented with the existing technologies to develop Electronic Travel Aids (ETAs) [6], many human factors have been overlooked [7].

Developing a promising low-cost interactive wearable de- vice is still an unmet need. In order to build such a wearable device, some technical challenges must be overcome. Of primary concern is the need to make the device easy to use and reliable. Ease of use refers to the immediate understanding of interaction with the wearable device, and reliability concerns safety levels and the expected response's effectiveness.

For this purpose, the design must overcome the following obstacles: 1) build the device as straightforward and simple as possible, 2) form a reliable interface, 3) overcome human errors, and 4) design the device to be power efficient.

We consider the need for a wearable device equipped with radio-frequency identification (RFID) to address these challenges. The communication is then performed through customized audio recordings on an onboard memory (database). All the elements are mounted on a glove. The proposed assistive glove acts as an interface for a fast connection to the environment, ensuring wireless communication between the tagged objects and the transponder on the glove.

This paper is organized as follows. Section II surveys related work, highlighting the benefits and limitations of available technologies. Section III represents the design and implementation of the proposed system. Section IV describes the hypotheses and experimental results and evaluates the reliability of the system. Section V discusses the findings, and Section VI concludes the paper.

## II. Related Work

People who live with blindness or severe visual impairment may use various technologies to cope with their daily activ- ities such as reading and navigation. The oldest travel aid, the white cane, is now augmented with capabilities such as obstacle's height and distance detection [8] and multimodal haptic feedback [9]. Others attached an embedded system to the conventional white cane and, by using electromagnetic sensors or magnetic coils, exhibited better performance [10], [11]. One of the most recent assistive devices proposed by A. Aladre'n *et al.* [12] is composed of an RGB-D camera (a depth sensor) that informs the user of nearby objects. Another camera-based indoor way-finding assistive method is based on 6-DOF pose estimation [13]. The device's form factor, however, resembles a conventional white cane. Both mentioned studies have a speech interface.

Many commercial solutions for the visually impaired focus on navigation and localization. Krishna [14] and Velázquez [15] introduced some outdoor tracking methods for navigating the visually impaired. These solutions are generally GPS-based but suffer from low accuracy in indoor environments.

Envision [16] is an assistive navigation system that utilizes a smartphone as a platform for GPS technology and a novel obstacle-detection method for navigation.

RFID technology has recently attracted attention in the research community, and utilizing it as localization and navigation proved to be a reliable method in indoor environments [17].

Dionisi *et al.* [18] developed a wearable object detection system, particularly for the visually impaired. The system searches for medication in the home environment using RFID detectors. It uses a vibrotactile communication method.

Another study conducted by Krishna and Balasubramanian [19] resulted in a wearable system for complete access in shopping environments. The comfortable strap-on glove consists of RFID wireless technology made it faster for identifying products. However, the system lacks independence since voiceovers are pre-recorded (implemented in a separate database). Also, a portable digital assistant (PDA) is required for communication via the Internet.

Similarly, Neela Harish [20] developed an experimental glove to use a Braille pad system as an interface.

NavCane [21] has a GPS module and provides both tactile and auditory feedback. Despite low cost and power, it is still in the form of a conventional white cane and might decrease the user's freedom of movement.

A study on the usability of speech-based interfaces [22] authenticates that the visually impaired appear to be more capable of processing spoken information than sighted subjects and verifies the importance of a speech-based interface.

Merler *et al.* [23] proposed an RFID-based device with a new multimedia database of groceries. The device's priority is accessibility, and people with normal vision can also use it. However, no information beyond what is included in the list is available. Also, users reported that holding the device in hand impedes the freedom of arm movements.

Tatsumi *et al.* [24] developed an RFID-based system for educational settings. Similarly, it has a voice version but requires a PDA for the database.

In another attempt [25], the team used RFID tags and traced them by autonomous robots for aiding the visually impaired to navigate. They deployed a self-organizing map, which is not feasible and requires robots and receivers. Therefore, they have to adjust the algorithm parameters for a simple task, especially to control the robot's navigation.

As for the recognition of objects, Cang Ye [26] recently put forward a robotic navigation aid with real-time detection of objects. Zhou *et al.* [7] implemented a smart object-detecting system that gives the user an interactive audible display about the proximity of nearby objects through a smartphone.

Ghiani *et al.* [27] investigated vibrotactile feedback to aid blind users with mobile guides. In this application, both sound and vibrotactile feedback were investigated. Since the database is located on a PDA, the user must carry a mobile device. As discussed in [28], the hands-free nature of technologies for the visually impaired should not be overlooked.

### III. Design Concepts and System Architecture

Wearable assistive devices are designed based on specific requirements, some of which are gathered in [29]. Our proposed system's key features include (1) safety, (2) simplicity, (3) reliability, and (4) having an independent onboard database. The designed version is primarily a prototype that is tested for the above characteristics.

The proposed system's primary function is to identify digital tags affixed to the nearby objects and initially manage to associate audio recording information to their unique IDs. This configuration allows the user to build an onboard database where all the necessary information about the target is stored. In case the same ID is further detected, the recorded message is recalled and played back to the user via a headphone.

According to Fig. 1, the system comprises two main components: (1) the tags, pre-located in specific locations or on objects, and (2) the wearable device, mounted on a glove.

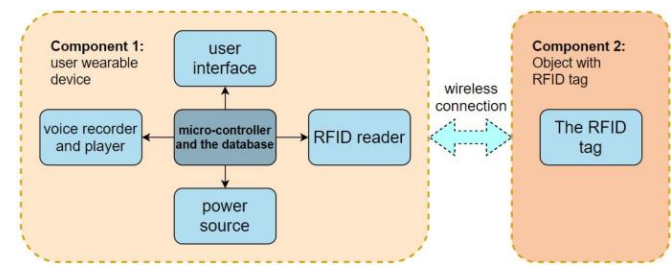

Fig. 1: The architecture of the system. The assistive glove comprises of an RFID transponder, the processing unit (microcontroller and database), the voice recorder (and player), the user interface, and a power source (battery) (left), and the RFID tags, pre-located in certain locations or on objects (right).

The wearable device consists of five main parts: the RFID transponder, the processing unit, the voice recorder (and player), the user interface, and the power source (Fig. 2). The elements of the proposed assistive glove are introduced in this section.

#### A. The RFID Transponder

The system's critical component is the RFID transponder that interacts with a set of tags with unique IDs. RFID tags are generally classified into two groups: active and passive. Active tags are mostly powered by embedded batteries. They emit stronger signals hence provide a more extensive detection range.

Passive tags are generally smaller and more affordable. They have an onboard capacitor that is charged when it receives a signal from a passive tag reader with the corresponding frequency. Once the tag is powered, it transfers its embedded ID code. In this work, we use passive tags with ISO 14443A standard and 13.56 MHz frequency.

on metal surfaces, other materials such as plastic, fabric, or wooden objects were chosen for the experiments.

Since the scanning range is within 5cm, the effective angle is within 60degrees. At 0° or 180°, the antenna has the maximum gain of about 5.5dBi [30]. As the angle expands, the reading latency increases. When there are multiple tags within a 5cm range, the reader detects the most direct one.

### B. The processing Unit

Because of its ease of use and low price, the prototype was first implemented on the Arduino UNO platform. The board is powered by an Atmega328 processor, which operates at 16MHz. However, after launching the system, the Arduino showed several weaknesses.

Firstly, the processor had a relatively low speed. Meanwhile, high speed for identifying tags and sending responses was one of this system's main requirements. Moreover, limited storage for audio files was one of the drawbacks of the Arduino.

Due to the board's incompatibility, external software had to convert audio files to another readable format. Plus, the lack of internal storage unit made it impossible to build a database.

All the mentioned disadvantages led to using a Raspberry Pi 3B. Unlike the Arduino UNO, Raspberry Pi's have no internal storage. Instead, they rely on removable microSD cards, allowing the user to save enough audio recordings. Being self-reliant has made this microcomputer suitable for building a customized database.

With a 1.4GHz 64-bit quad-core processor, this board is faster and more reliable than its predecessors and the Arduino UNO. This model comprises a 3.5mm audio out jack connected to a headphone for audio prompts. Also, having 4 USB
2.0 ports allows a connection to a microphone.

As shown in Fig. 2(a), the processing unit is mounted on the back of the glove connecting the RFID transponder and the voice recorder to the database.

### C. Voice Recorder

Once a new tag is detected (*i.e.*, the new ID does not exist in the database), the user receives an audio notification. The system is then ready to record the user's voice. To implement this configuration on the proposed platform, it was required to add a USB sound card adapter. This sound card is an intermediate module between the microphone and the processing unit (Fig. 2(a)).

A small push-button was implemented on the side of the index finger as the user interface. Once the new tag is detected, pressing this button would allow the user to record a 3 seconds audio message. The time duration can be adjusted based on the user's requirements. After the operation, it automatically returns to the detection mode. Moreover, pushing this button while scanning an already existing ID allows the user to delete the old recording and add a new one.

The microphone is connected to the voice recorder through the 3.5 mm audio jack. To record and store audio files, the PyAudio library was implemented. Although this library had limited functions, it could easily record and store audio for our application.

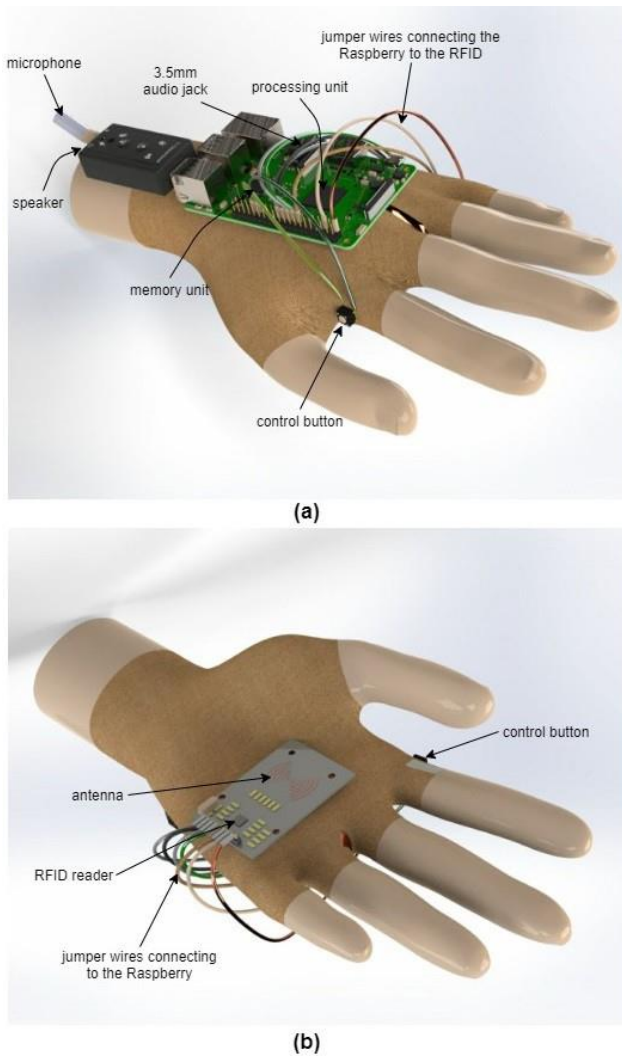

Fig. 2: The placement of the modules on the glove (left hand); (a) the processing unit (Raspberry Pi module), the voice recorder (USB channel sound card as well as the 3.5 mm headphone jack), placed on the back of the glove (battery and headphone not shown), (b) the transponder placed on the front of the glove.

We also selected a 13.56 MHz MF-RC522 module with low power, low cost, and an easy-to-use choice for the transponder. The transponder communicates with the processing unit via the serial peripheral interface (SPI) protocol with a maximum data rate of 10 Mbps. It is noted that this configuration also supports I2C and UART protocols.

To increase the detection range, the transponder is placed on the front (palm-side) of the hand (Fig. 2(b)). The module's dimension is 30 mm by 60 mm, and this placement does not limit the hand's range of motion. Since humans have more control over their hands, the performance is more precise when the sensor is on the front of the hand. The rest of the components were placed on the hand's back (dorsal side).

This configuration provides a detection range of roughly 5 cm. Hard 18 mm disk tags were used and affixed to the objects using double-sided tape. As the tags could not function

After recording each sound, the system automatically saves the audio file and associates it with the tag ID in the database. The flowchart in Fig. 3 summarizes the process.

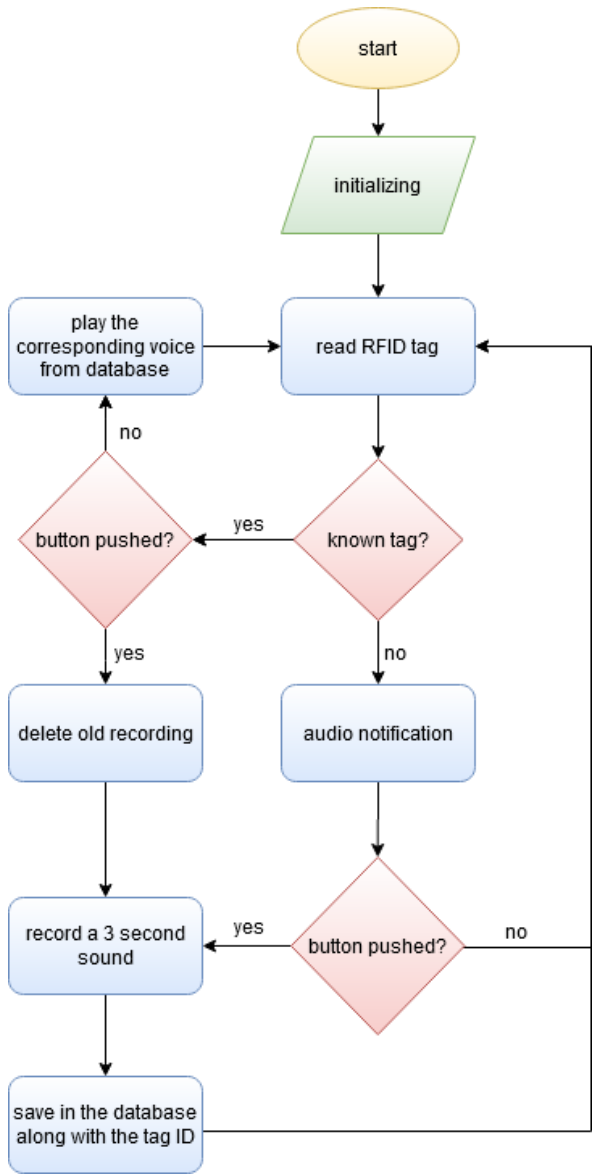

Fig. 3: The function of the proposed system: RFID tags affixed to the nearby objects are initially identified and associated with personalized audio recordings to gradually build an onboard database.

The glove was programmed to automatically remove the existing audio file and replace it whenever a new voice was recorded. This feature allows the user to use the glove more flexibly, replace the sound corresponding to each tag, and then assign it to a new object.

## IV. METHODS AND EXPERIMENTAL RESULTS

This section includes the experimental setup, the experi- ments, the data collection method, and the results.

### A. Participants

The usability and performance of the device were tested by performing indoor experiments. Seventeen healthy participants were included in this study. Participants' ages ranged from 18 to 30 years ($M$ = 23.6, $SD$ = 3.9), with 53% identified as female. The experiments were carried out randomly between 10 am to 6 pm for two weeks in a university laboratory. All the participants complied with the same predetermined protocol. Initially, each participant was asked to provide basic demographic information and sign an intent letter.

The participants were then asked to put on the glove and perform the tasks as instructed. A short introduction session was held for the participants. During this session, instructions on how to use the glove were given to them. Four experiments were performed in a controlled environment while the participants were blindfolded. In the end, they were given a questionnaire to complete.

### B. Experimental Setup

To evaluate the performance of the proposed system, various scenarios were considered, and four experimental setups were designed. The goal was to simulate the real experience of the visually impaired. Each experiment was exclusively designed to measure the effectiveness of the device based on different situations. All participants performed these tests in a random- ized order, and the number of each test merely indicates the name designated to it.

***Setup 1***: Eight objects with different shapes, colors, and materials were chosen. Eight RFID tags were attached to these items.

***Setup 2***: As shown in Fig. 4, the experimental setup consists of a box with three isodiametric holes. An RFID tag indicates each hole by a specific color. Nine circular disks with the same size as the holes were categorized into three color groups. Each circular disk had an RFID tag attached to it, indicating its color.

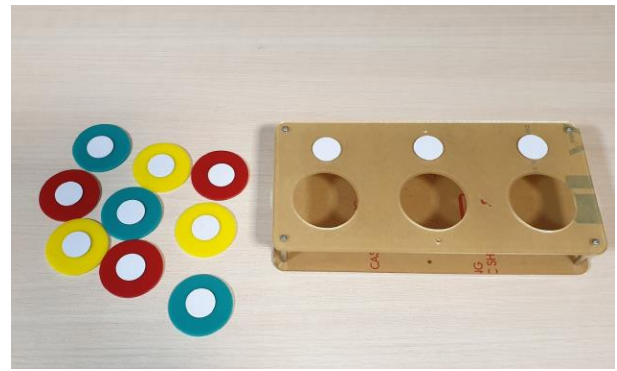

Fig. 4: Setup 2. Nine circular disks with different colors were used along with one box with three holes of the same size. Each circular disk and hole had an RFID tag attached to them.

***Setup 3***: The setup is shown in Fig. 5. It consists of a box with three holes yet with different shapes. An RFID tag is attached to each hole. Nine disks in the forms of 12, 14, and 16-slide polygons were also tagged separately.

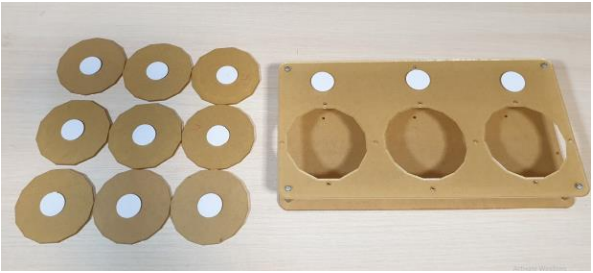

Fig. 5: Setup 3. This setup consists of a box with three holes in the form of 12, 14, and 16-slide polygons and 9 disks with the same shapes.

These parts are designed in a way that there was a possibility of error, and the person cannot perform the test by trial and error.

*Setup 4:* The table in Fig. 6 is divided into 8 regions. The regions are randomly tagged with numbers 1 to 8 on the side of the table. In order to consider human errors, more than one disk is placed on each section. The disks representing objects are also labeled with RFID tags by alphabet characters A, B, and C.

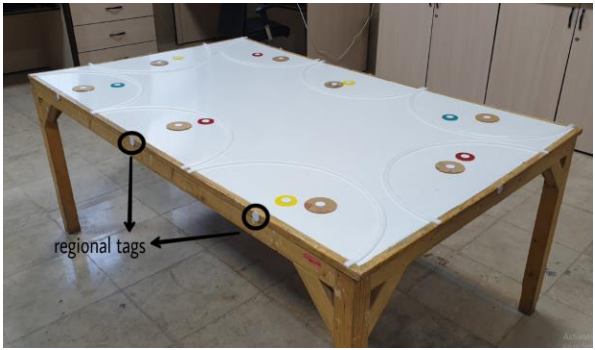

Fig. 6: Setup 4. A rectangular table was divided up into 8 sections, labeled with RFID tags by alphabet characters. There are two or three disks inside each region.

### C. Hypotheses and Experiments

Four hypotheses were formulated and analyzed using the IBM SPSS Statistics platform, version 23.0 (IBM Inc., NY). The results are presented according to the following experi- ments as well as the results of the survey based on the questionnaire.

*Test 1:* One of the proposed system applications is to enable the users to associate an audio message to an RFID tag. In this experiment, interactions with the system were tested.

The participants were instructed to define and characterize objects for the glove while blindfolded. They were supervised by an individual informing them about each object (Fig. 7).

All eight objects have a tag attached to them. Some tags are already associated with recordings in the database. If a pre-recorded audio already exists in the database, the saved audio recording is played. If the tag is unknown, the participant can enter new information by pressing the control button located on the glove.

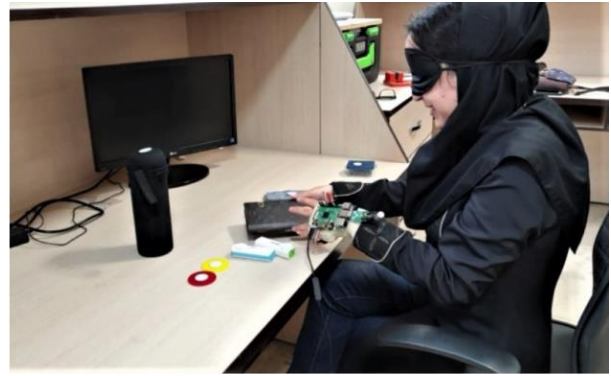

Fig. 7: Participants perform test 1 on eight different objects. The goal is to identify these objects, and in case of an unknown tag, they enter new voice recordings with the help of a supervisor.

The objective is to see whether participants can easily interact with the glove and put in information. Therefore, the success rate for all individuals was measured. Also, the time factor was recorded to determine if interaction with the glove becomes faster and more convenient with practice.

*Hypothesis 1:* The performance improves after using the glove several times.

This hypothesis seeks to show that after identifying each object, the performance gets better and faster since the user learns how to interact with the device.

*Result 1:* Completion rate is among the most fundamen- tal usability metrics. Based on the collected data, this task's success rate is measured by dividing the number of users who completed the task by the total number who attempted it.

In this case, 13 participants out of 17 completed all tasks correctly. Accordingly, the success rate can be measured as 76%. Three were 87.5%, and one person was 75% successful. The mean success rate was 96.32%.

In this test, the time taken to introduce the first to eighth object to the glove was recorded. The average time for attempt one to eight is $M = 70.17$, $SD = 12.22$ in seconds for first, $M = 57.70$, $SD = 12.09$ for second, $M = 50.70$, $SD = 8.60$ for third, $M = 43.76$, $SD = 8.48$ for forth, $M = 36.00$, $SD = 7.24$ for fifth, $M = 31.29$, $SD = 8.78$ for sixth, $M = 30.05$, $SD = 21.06$ for seventh and $M = 25.29$, $SD = 14.54$ for the last one.

A one-way repeated-measures ANOVA test was run to determine whether the time factor for identifying objects one to eight was different for the same group of 17 participants.

The Greenhouse-Geisser correction determined that the mean for objects differed statistically significantly in time ($F$ (3.438, 55.009) = 28.424, $P < 0.001$). Post hoc tests using the Bonferroni correction revealed that the time values reduce after several attempts.

Comparing the first objects' time values to the last three, we can see a statistical difference ($p < 0.001$). On the other hand, the last three attempts showed no difference ($p = 1.000$). This indicates that the time spent on identifying objects remains relatively stable after several attempts. The average time for the last three attempts is 27.87 seconds.

It can be concluded that with a 96.32% mean success rate, it takes approximately 28 seconds (after several attempts) for the user to identify and introduce objects to the system properly.

*Test 2*: One of the primary limitations of people with severe visual impairment is the comprehension of an object's visual features, such as color and fine patterns. One use of the glove is that, if needed, an object's visual characteristics can be introduced to it. In the second experiment, we investigated whether the glove is beneficial for recognizing colors.

For example, a blind person can identify a shirt, but they are unsure of the right color or design. Using the glove and only once with a sighted individual's help, they can identify the shirt, mark it, introduce it to the glove, and later independently find the desired shirt.

The participants were blindfolded and asked to put all the disks in corresponding holes. After the candidate performs the test, the overall time elapsed, and the disks' correct placements were recorded. The whole process was only conducted once by each participant.

   ***Hypothesis 2***: Participants can easily identify colors while blindfolded with a minimum number of errors.

   ***Result 2***: The participants were expected to put all nine disks in their correct place. The number of successful attempts for all individuals was 9 out of 9. Therefore, the success rate for this test was 100%.

Considering the objects were appropriately tagged with the one-time help of a caregiver, it can be said that a visually im- paired person can easily discern colors independently without error (mean of error = 0, N = 17).

Time values ($M = 114.0$ & $SD = 16.64$) were normally distributed ($p > 0.05$). Since the data concentration of M ± SD is within 68% (due to the empirical rule), it can be said that the system performs the same way for all individuals, regardless of gender, age, intelligence, and education.

*Test 3*: The visually impaired rely on tactile feedback in most daily activities. Unlike the previous test, in this test, objects are separable by the sense of touch.

In this experiment, the volunteers performed the test twice during the day. Once with the glove (w) and once without it (w/o). However, they performed the two tests in random order. The volunteers were supposed to put all 12 disks - visible in fig. 5 - in their dedicated place in the box. This task is expected to be carried out in less time and with fewer errors using the glove.

   ***Hypothesis 3***: Performance is more efficient with the glove than relying merely on the physical and tactile abilities (Alternative Hypothesis).

   ***Result 3***: To compare the two sets of data obtained from the test, we set up an equation to calculate a score for each data set. First, the accuracy percentage for all nine moves was measured in Eq. 1.

To calculate the accuracy, we subtracted the number of errors (e) from the correct moves (c). However, the value of a correct move was considered three times more than that of each error because completing the task was the objective of this test, and counting the errors was not of much concern. Further in the equation, nine is the number of moves.

$$Accuracy = \frac{(c - \frac{e}{3})}{9}\% \quad (1)$$

In Eq. 2, the final score is measured. In this equation, the time factor (t) was subtracted by the mean time value of all individuals' performance to normalize the time variable. This way, a value above the mean would be a positive, subtracted from the accuracy that decreases the score. Otherwise, a negative number that indicates less performance time would add to the final score. However, time was not of much value compared to accuracy. Accordingly, it was divided by three, which is an arbitrary figure proportional to the data.

$$score = Accuracy - \frac{t - \bar{t}}{3} \quad (2)$$

After removing the outliers, the data consists of 15 samples shown in Fig. 8. Applying the student's two-tailed paired t- test shows that using the glove resulted in higher accuracy and took less time.

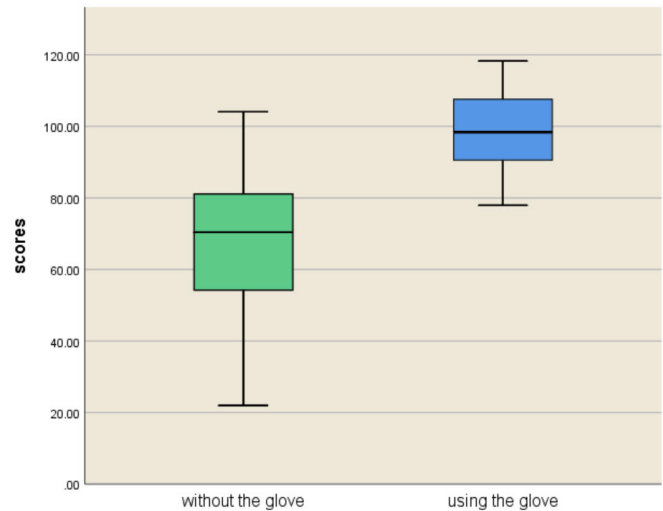

Fig. 8: The left box shows the scores obtained with tactile feedback and physical abilities, and the right box shows the scores obtained by the same group of subjects using the glove. The chart illustrates a significant difference between the two tests with higher scores for the group using the glove.

Table I shows the values of t-tests for three pairs of variables (score, time, and the number of errors). With a difference of 29.3≈ in the mean, the volunteers gained higher scores while performing the task using the glove than not using it. Therefore, it can be said that the performance is more efficient with the glove.

*Test 4*: The participants were instructed to locate a specific object (object A) from a randomly chosen region and then place it in a requested destination. To design the setup less biased, it was decided that all participants must turn around the table counterclockwise. Both the object and the destination were randomly chosen for each participant.

The parameters that were measured include (1) the total performance time $t_T$, (2) the amount of time taken to find the desired object $t_1$, (3) the number of section tags scanned to locate the origin n1, (4) the number of section tags scanned to locate the destination $n_2$.

TABLE I: T-test results for three variables with descriptive information.

| variable | | mean | std dev | t-value | p-value |
|---|---|---|---|---|---|
| score | w/o | 70.32 | 26.39 | -4.27 | 0.001 |
| | w | 99.62 | 10.70 | | |
| time | w/o | 249.0 | 99.45 | 5.24 | < 0.001 |
| | w | 106.8 | 30.62 | | |
| number of errors | w/o | 2.4 | 2.41 | 3.66 | 0.003 |
| | w | 0.06 | 0.25 | | |

*Hypothesis 4*: The glove's performance is similar for each individual regardless of the path taken. In order to compare the results, it was necessary to turn all the parameters mentioned above into one. For this purpose, a value was calculated by Eq. 3 and all these parameters were converted into a score.

$$score = \frac{t_T}{n_1 + n_2} + \frac{t_1}{n_1} \quad (3)$$

*Result 4*: The score data was found to be normally distributed ($p$ = 0.356). The coefficient of variation (CV) shows the extent of variability in relation to the mean. In this case, CV is 0.307, offering the data's precision. Therefore, with the success rate of 100% and the closeness of the scores ($M$ = 25.43, $SD$ = 7.82), it can be said that the glove's performance is similar regardless of the path taken.

*D. Reliability*

A system is reliable once the same sets of experiments (using the device) conducted on the same group of items at different times generates the same or similar results. Considering the glove as a sensor, we have to ensure it gives us the same results under the same conditions.

To measure the system's reliability, 8 RFID tags were attached to 8 objects, including one pair of socks, a box of chocolates, one yellow and one red circular disk, a blue and a green power bank, one notebook, and a bottle of water.

A one-way repeated-measures ANOVA was run between multiple tests (the task was tested 17 times). The P-value of 0.060 revealed that the time factor between multiple testings was not statistically significantly different ($F$ (4.91, 34.371) = 2.379), confirming the device's reliability.

In order to evaluate the user's experience, a questionnaire was designed. Following the work presented in [31], fifteen measuring items were ranked by candidates in the questionnaire.

The mentioned questionnaire consists of 5 sections. In the first part, volunteers enter their personal information and are asked questions about their age, education, and level of vision. The second part consists of 15 general questions about the device. The volunteers were asked to rate the device based on its ergonomics, comfort, personalization, simplicity, privacy, reliability, aesthetics, and wearability. The respondents could answer on a rating scale from disappointing (with a score of 1) to excellent (with a score of 5).

The third part contains eight technical questions regarding the device's speed, detection range, and sound clarity. The fourth part is a skip section, which includes five customized questions for the visually impaired and the blind, and the last section inquires a comparison of each previously performed task with and without the glove. The ratings for this section is scaled from impossible to very simply.

Regarding the general questions, the results show that 60% of the candidates found the device excellent in being wearable and reliable. Also, 53% of them found the glove very comfortable. Overall, 70% found the device very satisfactory in general. The device weighs around 92g, and the participants felt that it was very light on the hand.

The Cronbach's alpha reliability method was used to rep- resent the proportionality of a group of items measuring the same structure in the questionnaire. The reliability value for six items concerning the" effectiveness" factor is 0.782, and for" simplicity," with 6 statement items, it is 0.762. Both alpha values are above 0.5, indicating the high reliability of the questionnaire.

*E. Energy Efficiency*

The energy consumption of the system when the power is on includes two states: First, the sleep mode, which is when the RFID transponder is not in the vicinity of an object to identify and the only current consumption is the energy consumed by the single board Raspberry Pi 3, which is 400mA (2.0 Wh) in the inactive state.

Second, the active mode is when the RFID reader works and consumes over 32mA, with the maximum total USB peripheral current draw of the Raspberry Pi 3 computer board being 1200mA. Additional current consumption includes using the control button, recording and playing audio files, which brings out the total value up to 1400mA ($\approx$ 7W) in the active state.

At different time intervals, the energy consumption of the system varies. Considering the average 40% active and 60% sleep modes, the glove can last for 2.5 to 3 hours with a 2000mAh power supply.

Although energy efficiency was not the main focus of this work, it was revealed that the system is power-efficient enough for our application through experiments.

## V. CONCLUSION AND FUTURE WORK

We proposed design details and the experimental results of a wearable test prototype to identify objects and navigate through them. The device was designed specifically for people with visual impairment. To evaluate the system's performance, this system was tested and surveyed by 17 candidates and obtained acceptable results.

As demonstrated in the first test, the device's mean success rate in identifying objects is about 96%. On average, it takes the user 28 seconds to do a simple task (identify a tag and introduce a new object) with less than a 4% chance of failure. In the second test, it was proved that the visually impaired could discern colors using the glove. Besides, some tasks have

8attested better results using the glove than relying merely on physical abilities, as illustrated in the third test.

Unlike other tests where the participant was seated, in test 4, we showed that simple tasks such as relocating objects are also possible while moving.

The features of the proposed work can be listed as follows:

- This tool enables a new type of communication for visually impaired people to interact with the environment and objects around them.
- This glove is self-supporting and does not require devices and peripherals such as computers or the Internet (PDA).
- Due to its optimal and straightforward structure, this instrument will have a relatively low cost for the user and, therefore, can be easily accessible to the visually impaired community.
- This glove can be personalized and customized according to the user's needs.

Currently, the proposed device is in a prototype state. Despite some constraints, it has an expandable platform. To begin with, the USB sound card could be replaced by wireless headphones, and a smaller Raspberry version called" Raspberry Zero" could be used.

The coverage can be waterproof for more applications (such as in the kitchen). The dimensions of the glove can be reduced by using integrated electronic boards.